\documentstyle[epsf,prl,twocolumn,aps]{revtex} 
\tighten 
\begin{document} 
\draft

\title{Periodic  Orbit  Theory  of the  Transition  to Chaos in  Quantum  Wells}
\author{${\rm E.}^2$~Narimanov  and  A.~Douglas~Stone}  
\address{Applied  Physics, Yale
University, P.O.  Box 208284, New Haven CT 06520-8284} 
\date{November 14, 1995; revised April 16, 1996} 
\twocolumn[
\maketitle  \widetext   \vspace*{-1.0truecm}   
\begin{abstract}   
\begin{center}
\parbox{14cm}{  An  analytic  theory  is  developed  for the  density  of states
oscillations  in quantum wells in a magnetic  field which is tilted with respect
to the  barrier  planes.  The main  oscillations  are  found  to come  from  the
simplest  one or  two-bounce  periodic  orbits.  We  calculate  their period and
stability  analytically  and  find  an  infinite  sequence  of  destabilizations
followed by restabilizations as the chaos parameter  increases.  This phenomenon
explains  the  re-entrant  frequency-doubling  of the  density  of states  peaks
observed in recent magnetotunneling  experiments.}  \end{center}  \end{abstract}
\pacs{ \hspace{1.9cm} PACS numbers:  05.45.+b, 72.15.Gd, 73.20.Dx} ] \narrowtext

Most of our intuition  about the level  structure of quantum  systems comes from
the  consideration  of  hamiltonians  with high symmetry for which the classical
motion  is  integrable  and  hence  the  Schr\"odinger  equation  is  separable.
Symmetry-breaking terms are treated by perturbation theory.  This approach fails
when the  symmetry-breaking  terms  become  too  large and  typically  numerical
solution is employed.  An alternative  approach which has been used successfully
in atomic physics during the past decade is to use periodic orbit theory for the
density of states (DOS) of non-integrable systems.  Until recently this approach
has had few applications in condensed matter physics and its power is not widely
appreciated.  Recently  however  Fromhold  et al.  \cite{Fromhold}  studied  the
tunneling I-V  characteristic  through a wide quantum well in a strong  magnetic
magnetic  field  oriented at an angle $\theta$ to the tunnel  barriers.  As they
varied the  voltage at fixed  magnetic  field ($B = 11 {\rm T}$) they found that
resonant   tunneling  peaks   (corresponding  to  the  sub-band   thresholds  at
$\theta=0$) would abruptly change in frequency.  Simple estimates show that this
behavior occurs well outside the regime in which sub-band  mixing can be treated
by  perturbation  theory.  Modelling  of the  classical  dynamics  in this  case
indicated  that the motion was  completely  chaotic and the new structure in the
tunneling  density of states (DOS) was explained by the  appearance  of new {\it
unstable}  periodic orbits via Gutzwiller's Trace formula  \cite{gutzwiller}.  A
very important  later  experiment  looked at the I-V peaks in the entire (plane)
parameter  space of magnetic field and voltage for relatively  small tilt angles
\cite{Muller}.  At small tilt angles the  frequency-doubling  behavior was found
at much lower magnetic field ($B\approx 5 {\rm T}$) but was ``re-entrant'', i.e.
would  disappear  at yet higher  fields  (Fig.1).  At higher  tilt angles a more
complex pattern of peak-doubling  {\it and} tripling is observed  \cite{Muller}.
This  experiment  was of particular  interest  because it mapped the  transition
regime to chaos in which stable and unstable  periodic  orbits  exist and evolve
according to the general  constraints  governing  the  transition  to chaos.  No
previous experiments of this type exist in condensed matter physics.

Stimulated by this  experiment  Shepelyansky  and Stone  \cite{ss}  modelled its
non-linear  dynamics and showed that the system  underwent a KAM  transition  to
chaos   described  by  a  2D   effective   map  similar  to  the  standard   map
\cite{chirikov}  in which the degree of chaos was controlled by a  dimensionless
parameter $\beta = 2 v_0 B/E \sim B/\sqrt{V}$ (where $\varepsilon_0 = m^*v_0^2/2
\approx eV$ is the total injection  energy).  Hence at fixed tilt angle $\theta$
the degree of chaos is  constant  along a  parabola  $V \sim  B^2/\beta^2$,  but
increases  with  increasing  $B$  at  fixed  $V$   \cite{effmass}.  The  initial
frequency-doubling  island occurs well before the dynamics becomes fully chaotic
and ref.  \cite{ss}  noted  that its  onset  appeared  to be  associated  with a
period-doubling   bifurcation   of  a  {\it  stable}   period-one   orbit  which
surprisingly {\it  restabilizes} for even higher $\beta$ (chaos  parameter).  In
this work we  develop  for the first  time an  analytic  theory of the  periodic
orbits  controlling the main DOS oscillations in the transition region to chaos.
The orbits  responsible  for the  peak-doubling  at small  tilt  angles are {\it
different} from those previously considered  \cite{Fromhold} and account for the
re-entrant  magnetotunneling  behavior at small $\theta$ both  qualitatively and
quantitatively.  At higher tilt angles these orbits remain  important but others
previously considered \cite{Fromhold} make significant contributions as well.

The non-linear  conductance of the well is  proportional to the local DOS of the
well in the vicinity of the left (emitter)  barrier.  Electrons  tunnelling into
the well at high voltages gain kinetic  energy as they  accelerate  in the field
and collide with the collector barrier.  Over several collisions in the well the
electron  loses this energy by optic phonon  emission.  Therefore  the tunneling
resonances  are  substantially  broadened and only are sensitive to structure in
the DOS on energy scales $ >  \hbar/T_{opt}  \sim 5 meV$.  The Gutzwiller  Trace
formula relates the contribution to the DOS which is oscillatory  with energy to
a sum of  contributions  from each periodic  orbit (PO)  \cite{gutzwiller};  the
optic  phonon  broadening  implies  that  only the few  shortest  PO's will give
resolvable  structure  in the  experiment.  The Trace  formula also implies that
stable PO's typically give larger  contributions  than do unstable, so that in a
mainly  chaotic  phase  space the  biggest  DOS  oscillation  comes from the few
remaining stable PO's.

The  origin  of chaos in the  tilted  well is as  follows.  At zero  tilt  angle
($\theta =0$) the acceleration  along the electric field ${\bf E}=E \hat{z}$ and
the transverse  cyclotron  motion decouple and are  integrable.  Collisions with
the barriers ($v_z \to -v_z$) do not transfer  energy  between the cyclotron and
longitudinal  motion.  Once  the B field  is  tilted  (${\bf  B}= B \cos  \theta
\hat{z} + \sin \theta  \hat{y}$) the electron  executes  cyclotron motion around
the $\hat{ {\bf B}}$  direction, with a superimposed  drift velocity $v_d= (E/B)
\sin \theta \hat{x}$, and accelerates along $\hat{{\bf B}}$ due to the component
${\bf  E}  \cdot  {\hat  {\bf  B}} = E \cos  (\theta)$.  This  motion  is  still
integrable.  However now  collisions  with the barriers in general  {\it do} mix
the cyclotron and longitudinal energies  $\varepsilon_c,\varepsilon_L$  and make
the total dynamics  non-integrable.  The amount of energy exchange between those
two degrees of freedom  $\Delta  \varepsilon $ depends  sensitively  on the {\it
phase} of the  cyclotron  rotation at impact; for example if the velocity  falls
precisely in the $x-z$ plane  $\Delta  \varepsilon=0$.  Since the rate of change
of the phase is  $\omega_c=eB/m^*$,  we find  \cite{ss}  increasing  chaos  with
increasing  $B$.  Thus the  restabilization  of period-one  orbits at higher $B$
requires explanation.

With the choice ${\bf A} = (- B y \cos(\theta) + B z  \sin(\theta),  0, 0)$, the
system can be described by an effective  Hamiltonian with two degrees of freedom
\cite{ss,Muller}:

\begin{eqnarray}  H_{eff} & = & {{p_y^2 +  p_z^2}\over{2m}}  + {{e^2 B^2}\over{2
m}} (y \cos(\theta) -  z\sin(\theta))^2 - eE z \nonumber \\ &+& U(z) + U(-z - d)
\label{eq:Ham2} \end{eqnarray} \noindent

where $U(z<0) = 0, \ U(z>0) = \infty $.  If  $\varepsilon_0  > eV$, the electron
can retain enough  longitudinal  energy on collision at $z = 0$ to collide again
with the emitter barrier.  But if $\varepsilon_i  \equiv  \varepsilon_0 - eV \ll
eV$, (as is the case for the experiments of refs.  \cite{Fromhold,Muller}), then
$|\Delta  \varepsilon|$ is much larger at the collector  barrier $(z=0)$ than at
the  emitter  barrier.  Therefore  the most  salient  features of the  classical
dynamics can be described by the``1-barrier model'' (1BM) \cite{ss} in which the
emitter barrier $U(-z-d)$ is omitted in Eq.  (\ref{eq:Ham2})  above  \cite{1BM}.
The  periodic  orbit  structure  of  the  1BM  depends  only  on  $\beta,\theta$
\cite{ss}.  This  can be seen  directly  from  Eq.  (\ref{eq:Ham2}):  if time is
measured in units of  $\omega_c^{-1}$  and distance in units $E/B  \omega_c$ the
rescaled 1BM Hamiltonian depends only on $\beta,\theta$.

For $\theta = 0$ the system is integrable and {\it all} the periodic  orbits are
of two types.  First, degenerate families  corresponding to $\omega_c/\omega_L =
p/q$  where   $p,q$  are   integers.  Here   $\omega_L=   \pi  e   E/\sqrt{2   m
\varepsilon_L}$  is the  frequency of the  periodic  motion in the  longitudinal
direction  $\hat{E}=\hat{z}$ of the uniformly  accelerated electron bouncing off
the collector barrier.  Note that this frequency {\it decreases} with increasing
longitudinal energy as $1/\sqrt{\varepsilon_L}$  to its minimum value $\omega_L=
2 \pi  \omega_c/\beta$  when  all the  energy  is  longitudinal  ($\varepsilon_L
=\varepsilon_0)$.  Below this energy  (determined  by  $\beta_{min}=p/q$)  these
{\it helical}  orbits (HO) cease to exist.  However there is one {\it  isolated}
periodic orbit which exists for all energies.  This {\it traversing  orbit} (TO)
has no transverse energy $(\varepsilon_L=\varepsilon_0)$, its period is $T_0=2 m
v_0/eE=\beta/\omega_c$.  It is is distinct  from all  helical  orbits  except if
$\omega_c  T_0=\beta=2\pi  p$ ($p$ integer) when an  infinitesimal  increment of
transverse  energy generates a one-bounce HO of period $T_0$.  The TO determines
the  sub-band   energy   spacings  of  the  triangular   well  by   (integrable)
semiclassical  quantization, $ \delta E= \hbar/T_0$ and will be relevant for the
DOS oscillations at $\theta \neq 0$.

First, all one-bounce  orbits for $\theta \neq 0$ must be  ``non-mixing'',  i.e.
there can be no exchange of $\varepsilon_L$ and $\varepsilon_c$ at the bounce as
these are separately  conserved  between  bounces and the orbit is periodic.  It
follows that all such orbits must collide with with  $v_y=0$,  since this is the
condition for zero energy  exchange.  This  geometric  property of {\it all} the
one-bounce  orbits  allows us to to derive  their  periods,  $T$,  analytically.
Consider such an orbit in the frame of reference  moving with the drift velocity
$v_d = E  \sin(\theta)/B$.  In this  frame  the  orbit  projected  on the  plane
perpendicular  to  $\hat{B}$  is a portion of a circle  with a chord  removed of
angular size $2 \phi_0 = 2  \arcsin(v_d\omega_c  T / 2 v_c) $,where  $\phi_0$ is
the  cyclotron  phase  just  after the  collision,  and  $v_c$ is the  cyclotron
velocity.  Therefore, $v_c = (v_d\omega_c T / 2) \sin((2 \pi - \omega_c T)/2 )$.
Since the motion along $\hat{B}$ between collisions is uniform  acceleration, at
the barrier  $|{\bf v} \cdot  \hat{\bf B}| =  eE\cos(\theta)T/2m$.  Using energy
conservation we obtain (for arbitrary $\theta$)

\begin{eqnarray}  { {(\beta/2)^2 - (\omega_c T/2)^2} \over { \left(1 - (\omega_c
T/2)   \cot(\omega_c   T/2)\right)^2}}   =   \sin^2(\theta).   \label{eq:period}
\end{eqnarray}

The  solutions  of Eq.  (\ref{eq:period})  are shown in Fig.  2.  Note  that for
$\beta < 2\pi$ there  exists only one orbit with  period  $\omega_c  T_0 \approx
\beta$ as for the TO of the  untilted  system,  and two more  orbits  appear  at
thresholds given by $\beta^t_p  \approx 2\pi p ,  \;\;p=1,2,\ldots\infty$.  Thus
we have an infinite  set of {\it  tangent  bifurcations}  \cite{Lichtenberg}  in
which two new orbits (one stable and one unstable) appear; for small $\theta$ we
find from Eq.  (\ref{eq:period}) $\beta^t_p = 2 \pi p ( 1 + (\theta/\pi p)^{1/3}
+ O( \theta^{4/3}))$.  It follows from Eq.  (\ref{eq:period}) that for arbitrary
$\beta$  there  exist $2 N$  (with  $N = {\rm  integer}(\beta/\beta_N^t)$)  such
orbits whose periods are approximately an integer multiple of $2  \pi/\omega_c$,
and one orbit of longest  period, $T_0$, which  approaches the period  $\omega_c
T_0=\beta$ of the  unperturbed TO.  This orbit is a recognizable  deformation of
the  TO at  $\theta=0$(see  inset,  Fig.  2a),  whereas  the  other  orbits  are
deformations of the $\theta=0$  helical orbits of period $\omega_c T = 2 \pi p$.
The  asymptotes  at $\omega_c T = 2\pi p$ arise  because for $\theta \neq 0$ the
period $T =2 \pi p/\omega_c$ is not a solution of Eq.  (\ref{eq:period}) for any
$\beta$.  If it were a  solution  then in the drift  frame the orbit  would be a
full circle, which is impossible due to the collision.  The helical  orbits with
$\omega_c T_0 \ll \beta$ have substantial  cyclotron energy and typically remain
far from the  emitter  barrier  (see  inset, Fig.  2a); hence the TO is the only
period-one orbit important for the tunneling DOS and it defines the ``sub-band''
spacing of the tilted well \cite{Muller}.

The  stability  of these  periodic  orbits is  described  by a monodromy  matrix
\cite{comment1}  $M$; the orbit is unstable if $|{\rm  Tr}(M)|>2$.  For 1-bounce
orbits we find

\begin{eqnarray}  {\rm Tr}(M_1) & = &  4\cos^4(\theta)  \left(  \tan^2(\theta) +
(\omega_c   T/2)\cot(\omega_c   T/2)   \right)   \nonumber   \\   &   \times   &
\left(tan^2(\theta)  + sin(\omega_c  T)/(\omega_c T) \right) - 2 \label{eq:trm1}
\end{eqnarray}

The results (plotted in Fig.  2) show that for small tilt angles and $\beta \leq
4\pi$ (the  experimental  parameter  range) the TO goes  unstable  only in small
intervals of $\beta$ around $\beta = \pi, 3\pi, \ldots$.  This can be understood
by continuity.  One can easily show that at $\theta=0$ the TO is only marginally
stable ($ |{\rm Tr}(M)|=2$) at $\beta=\omega_c  T_0= \pi p , \; \; p=1,2 \ldots$
and is otherwise stable.  Since the transition to chaos is smooth \cite{ss} with
increasing  $\theta$,  instability  can only appear for $\theta  \ll 1$ in small
intervals  around  $\beta = \pi p$.  At $\beta =  \beta_p^{D}  \approx \pi (2p -
1)(1 + 3\theta^2  /2\pi^2(p-1/2)^2 $ the instability  occurs by  period-doubling
bifurcation and a rapid restabilization  occurs at $\beta_p^R \approx \pi (2 p -
1) (1 + \theta^2(2 + 1/\pi^2  (p-1/2)^2)$.  At $\beta  \approx 2\pi p $ as noted
above a stable and  unstable  period-one  orbit  appear so that {\it two} stable
period-one  orbits coexist; the one with the longer period is the TO which hence
``jumps'' in period.  (The  discontinuity in the peak data near $B\approx 10{\rm
T},  V\approx  0.4{\rm  V}$ in Fig.  1  occures  right at $\beta =  \beta^t_1$.)
After this jump the helical orbits rapidly destabilize with increasing  $\beta$.
For large $\theta$ or $\beta$ the TO never  restabilizes  after its  bifurcation
near $\beta \approx  \pi,3\pi,\ldots$, but it {\it always} reappears as a stable
orbit  at the  tangent  bifurcation  near  $\beta  \approx  2 \pi p$.  Thus  for
arbitrary $\theta$ the TO undergoes an infinite sequence of destabilizations and
restabilizations  and stable period-one  islands exist at arbitrarily high chaos
parameter.

How does this explain the re-entrant  behavior of the experimental  DOS?  As the
magnetic  field is increased  $\beta$ first passes through  $\beta = \beta_1^{d}
\approx  \pi$ and the TO  bifurcates:  for some  interval  of B there  exists  a
stable period-two orbit ($T_2 \approx 2T_0$).  The Trace formula implies that if
a stable  period-two  orbit  appears  there  should be a large  increase  in the
amplitude  of the DOS  oscillation  at  $\hbar/2T_0$  (see Fig.  3)  giving  the
period-halving.  Eventually the period-two orbit destabilizes by period-doubling
bifurcation,  its  contribution to the DOS is lost, and there is a return to the
original  sub-band  spacing  determined  by the TO  (see  Fig.  1).  However  in
contrast to simple  area-preserving 2D maps of the Chirikov type \cite{chirikov}
the  increase  in chaos is not  monotonic;  a new  stable TO appears by  tangent
bifurcation at yet higher B and its period-doubling bifurcation defines the next
interval of peak-doubling as seen in Fig.1.

Although the 1BM explains many of the puzzling qualitative  features of the data
of ref.  \cite{Muller},  for further insight and quantitative  comparison we now
consider  the  two-barrier  model,  Eq.  (\ref{eq:Ham2}).  The  PO's  in the 2BM
depend not only on $\beta,  \theta$, but also on  $\gamma=\varepsilon_0/eV$.  We
calculate  $\gamma  \approx 1.1 - 1.2$ for the experiment of ref.  \cite{Muller}
based on the  standard  model  for the  emitter  state  \cite{Ando}.  Using  the
non-mixing  property of the TO we can again  calculate  its period and stability
analytically  \cite{Evgeni} and (assuming  constant $\gamma$) we plot the result
in Fig.  2b vs.  $\beta$.  Note  that in the  2BM for  $\theta=0$  the TO  obeys
$\omega_c T_0 = \beta (1 -  \sqrt{1-1/\gamma})$  (not  $\beta$) and the behavior
for small $\theta$  approaches  this line except near $\beta = 2\pi p$ where the
helical  period-one orbits appear.  These orbits have too little energy to reach
the emitter barrier and hence are exactly as in the 1BM.  The TO also has a very
similar  geometry  as in the 1BM except of course for most  $\beta$ it  collides
with both  barriers.  However a significant  difference  arises in its stability
properties.  The  order  of  the  bifurcation  and  restabilization   points  is
interchanged  as a function of $\beta$ (or B) (see  insets,  Fig.  2a,2b).  This
implies that the period-doubling region occurs for $\beta < \beta^d$ (smaller B)
and not for $\beta > \beta^d$ as in the 1BM.  This  difference can be understood
by noting that the period-two  orbit requires  substantial  $\varepsilon_c$  and
hence  occurs only above a  threshold  energy,  $\varepsilon_2$,  determined  by
$T_0(\varepsilon_2,B) \approx \pi/\omega_c$.  In contrast to the 1BM, in the 2BM
$T_0$ is a {\it decreasing}  function of energy since a higher energy  collision
with the emitter  barrier  just returns the  electron to the  collector  sooner.
Thus the high-energy  region in which a stable  period-two  orbit exists, occurs
when  $\omega_c  T_0 < \pi$, i.e.  for smaller  values of $\beta$.  Denoting the
point of  bifurcation by $\beta_2$,  the  restabilization  occurs for $\beta_R <
\beta_2$.  At  $\beta_3 < \beta_R$  the  period-two  orbit  destabilizes  and at
$\beta_4 < \beta_3$ it disappears  in an (inverse)  tangent  bifurcation  with a
period-two  orbit which doesn't  reach the emitter.  Thus we identify  $\beta_4$
with the onset of  period-halving in the tunneling  oscillations in Fig.  1, and
$\beta_2$ with the return to period-one  oscillations.  The same sequence occurs
again at higher  $\beta=\beta_2',\beta_4'$.  These peak-doubling  boundaries are
calculated analytically from the 2BM using the estimated experimental parameters
(with no fitting) and are found to agree rather well with experiment (Fig.1).

The peak  behavior  at higher  tilt  angles  is more  complicated,  as the TO is
unstable  over a large  interval of $\beta$ and the  contribution  of it and its
offspring to the tunneling DOS needs to be compared with other unstable  orbits.
Comparison   to  the   data  of  ref.  \cite{Muller}   at   $38^{\circ}$   shows
peak-doubling  and tripling  regions in very good agreement with the bifurcation
and  trifurcation  of the TO and a  re-entrant  single-peak  region at high B in
reasonable   agreement  with  the   restabilization  of  the  TO  after  tangent
bifurcation  \cite{Evgeni}.  Other  regions  of  multiple  peaks  appear  to  be
associated  with  previously  identified  orbits  \cite{Fromhold}  which are not
bifurcations or  trifurcations  of the TO but whose interval of existence can be
understood in a similar framework \cite{Evgeni}.

In Fig.  3 we confirm  that the DOS for the tilted well does show  peak-doubling
in the interval  predicted by our theory.  Here we show a numerical  calculation
of  the  oscillatory  part  of  the  DOS  appropriately  scaled  using  Eq.  (1)
\cite{Evgeni}  for  $\theta=20^{\circ}$.  As the energy  increases  through  the
classical  bifurcation  point the DOS  oscillations  clearly pick up a secondary
maximum  corresponding  to the  period-halving,  but at even higher energies the
secondary  maxima disappear at the predicted  restabilization  point for the TO.
Note in Fig.  1 that the  secondary  maxima are not observed at lower and higher
$B,V$ where the  classical  theory  implies a stable  period-two  orbit  exists.
However at lower fields the period $T_2 \sim 1/\omega_c$ increases, so the optic
phonon  damping $\sim  \exp(-T_2/\tau_{opt})$  removes the  contribution  of the
period-2 orbit from the tunneling  spectra.  Hence  secondary  maxima die out in
the  period-two  layer at low fields  (see  lower  curve of Fig.  3).  At higher
$B,V$ we propose a different mechanism suppresses the peak-doubling.  Typically,
the period-two orbit has a substantial amount of cyclotron energy $\varepsilon_c
\sim  \varepsilon_0  - eV = (\gamma  -1)eV$, which is greater then the cyclotron
energy in the emitter state ($\sim \hbar  \omega_c$).  Tunneling to the period-2
orbit therefore  requires energy transfer from longitudinal to cyclotron motion.
The amount of this  energy  transfer  increases  with  increase  of $B$ and $V$,
leading to a decrease in the tunneling  matrix  element for this process at high
fields \cite{Evgeni}.

We thank  Greg  Boebinger,  Harsh  Mathur  and Dima  Shepelyansky  for  valuable
discussions, and G.B.  for access to unpublished  data.  This work was partially
supported by NSF Grant DMR-9215065 and by a President of Russia Fellowship.

\begin{figure}[bt] \epsfbox[0 300 300 580]{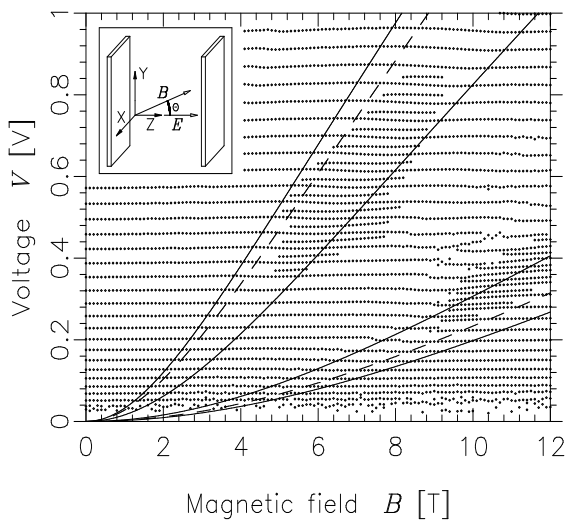} \caption{Resonant tunneling
peak locations for $\theta=11^{\circ}$  from [3].  Pairs of solid curves enclose
the regions of existence of period-two  orbit for 2BM.  Dashed line  corresponds
to the  destabilization  of the period-2 orbit.  Variation of the effective mass
with energy has been taken into account.  } \end{figure}

\begin{figure}  [bt] \epsfbox[100 500 900 1000]{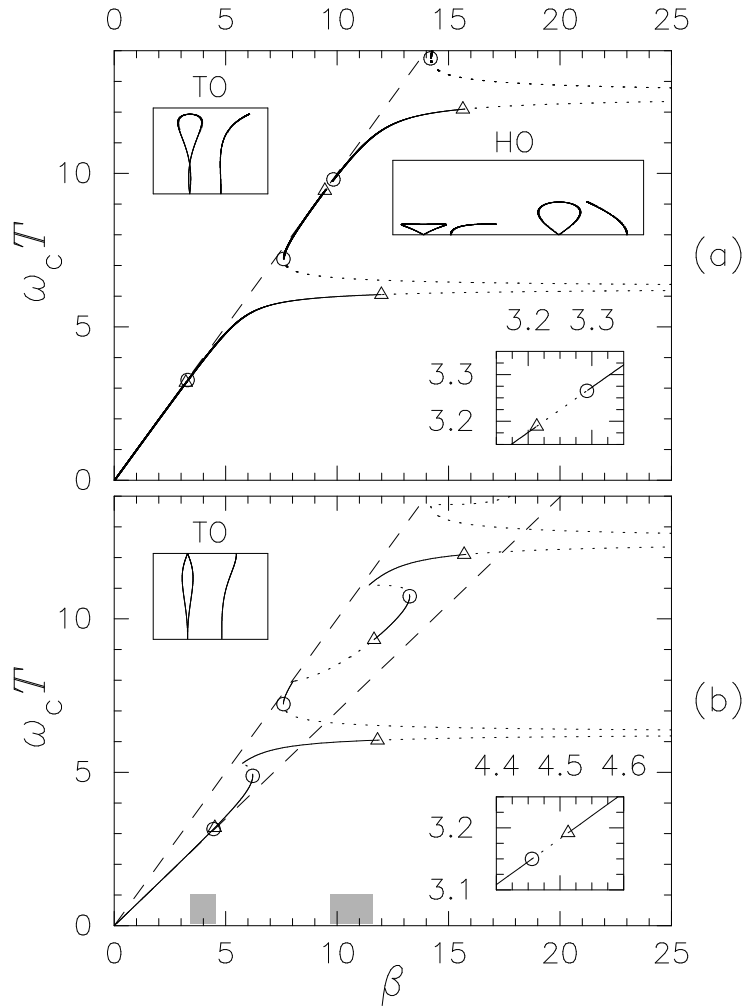}  \caption{\label{fig2}
(a) Periods of one-bounce  orbits vs.  $\beta=2(v_0B/E)$  for one-barrier  model
with  $\theta=11^{\circ}$.  Solid lines denote stable  orbit,  dotted  unstable.
Dashed  line  is  $\theta=0  $  result  for  TO.  Triangles  =   period-doubling
bifurcations,   circles  =  restabilizations.  Insets:  (top  left):  Traversing
orbit (TO) for $\beta=10$  projected in y-z (left) and x-y (right)  plane.  (top
right) same for helical orbits (HO).  (Lower right):  blow-up near  $\beta=\pi$.
(b) Same for two-barrier  model with  $\varepsilon/eV=\gamma=1.1$.  Lower dashed
line is  $\theta=0$  result for 2BM, upper for 1BM.  Grey-scale  regions  denote
intervals of existence of period-two orbit which is born in tangent  bifurcation
and disappears in a backwards period-doubling bifurcation.  } \end{figure}

\begin{figure}  [bt]  \epsfbox[0  300  300  500]{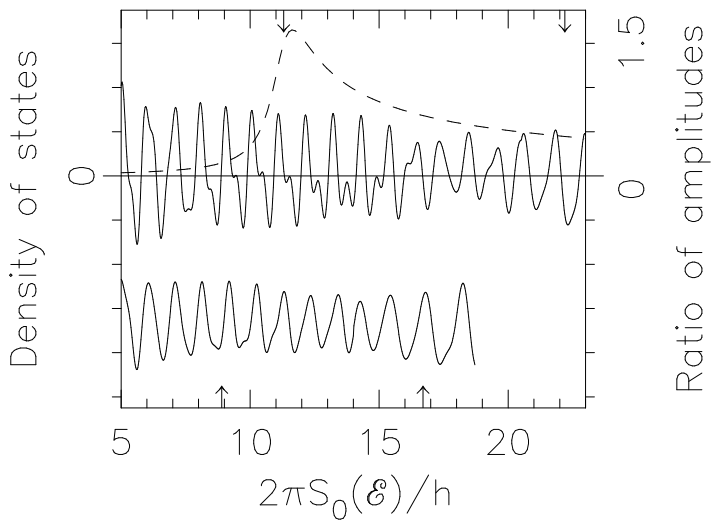}  \caption{\label{fig3}
Oscillatory    contribution   to   DOS   calculated   numerically   vs.   action
$S_0(\varepsilon_0)/\hbar$ of unperturbed TO.  Note onset of secondary maxima at
classical   bifurcation  energy  (1st  arrow)  and  disappearance  at  classical
restabilization  energy (2nd arrow).  (Bottom Trace, $B=3.53 {\rm T}, V=0.2 {\rm
V}, \theta = 20^0$ ):  stable  period-two  interval has decreased and shifted to
lower energies and secondary  maxima are no longer  apparent (see text).  Dashed
line is ratio of the period-two to period-one contributions to the DOS for $ B=5
{\rm T},  V=0.4  {\rm V},  \theta = 20^0$  from  Eqs.  (2),(3)  using  the Trace
formula   [2].  Level-broadening   is   $\hbar/T_{opt}   =  6.6  {\rm  meV}$.  }
\end{figure}


\begin{references}  \bibitem{Fromhold}  T.M.Fromhold  {\it et al.},  Phys.  Rev.
Lett.  {\bf 72}, 2608 (1994).  \bibitem{gutzwiller} M.C.Gutzwiller {\it Chaos in
Classical   and   Quantum   Mechanics}   (Springer-Verlag,   New   York,   1990)
\bibitem{Muller}  G.  Muller  {\it et al.},  Phys.  Rev.  Lett.  {\bf 75},  2875
(1995);  G.S.  Boebinger  {\it  et  al.},  Proc.  EP2DS-IX,  Surf.  Sci  (to  be
published).  \bibitem{ss}  D.L.Shepelyansky  and  A.D.Stone,  Phys.  Rev.  Lett.
{\bf 74}, 2098 (1995)  \bibitem{chirikov}  B.V.Chirikov,  Phys.  Rep.  {\bf 52},
263 (1979)  \bibitem{effmass}  The  parabola is flattened at high voltage due to
the  increase of the  effective  mass away from the band  minimum (see Fig.  1).
\bibitem{1BM}  With a slightly  different  doping profile the bottom of the well
could be raised above the emitter  state at zero bias  allowing an  experimental
study of the pure 1BM.  \bibitem{Lichtenberg} A.J.Lichtenberg and M.A.Lieberman,
{\it Regular and Chaotic dynamics} (Springer-Verlag, New York, 1994)

\bibitem{comment1}  The stability matrix of 1-bounce  periodic orbits is carried
out for the  effective 2D  Hamiltonian  (\ref{eq:Ham2}).  The full 3D orbits are
marginally  stable with respect to  translations in $x$; in the quantum DOS this
leads to the  familiar  Landau  level  degeneracy  which is not  relevant  here.
\bibitem{Ando}  T.~Ando  {\it et al},  Rev.  Mod.  Phys.,  {\bf 54}, 437  (1982)
\bibitem{Evgeni} E.  Narimanov and A.  D.  Stone, unpublished.  \end{references}
\end{document}